\documentclass{epl}

\usepackage{amsmath}
\usepackage{graphicx}
\usepackage{latexsym}
\usepackage{subfigure}

\newcommand{\bu}{\boldsymbol{u}}
\newcommand{\bv}{\boldsymbol{v}}
\newcommand{\bS}{\boldsymbol{S}}
\newcommand{\btau}{\boldsymbol{\tau}}

\newcommand{\bxi}{\boldsymbol{\xi}}
\newcommand{\be}{\boldsymbol{\eta}}

\title{Statistical Mechanics of Linear Compression 
Codes in Network Communication}
\author{Tatsuto Murayama}
\institute{
  Department of Computational Intelligence and Systems Science, Tokyo Institute of Technology - Yokohama 2268502, Japan
}


\begin{document}

\maketitle

\begin{abstract}
We analyze the performance of a linear code used for 
a data compression of Slepian-Wolf type. 
In our framework, two correlated data are separately 
compressed into codewords employing Gallager-type codes
and casted into a communication network through two independent 
input terminals. At the output terminal, 
the received codewords are jointly decoded 
by a practical algorithm based on 
the Thouless-Anderson-Palmer approach.
Our analysis shows that the achievable rate 
region presented in the data compression theorem by Slepian 
and Wolf is described as first-order phase transitions among 
several phases. 
The typical performance of the practical decoder is also well 
evaluated by the replica method.
\end{abstract}

The ever increasing information transmission in the modern world is 
based on network communications. While a lot of cutting edge technologies 
have been developed to realize the comfortable communication, 
few techniques are designed for network-based data transmission. 
It is quite strange that even the up-to-date information 
technologies are based on point-to-point protocols, 
although the global computer network called `the Internet' 
already exists. Therefore, now is the time that we focus 
on the multi terminal communication techniques.

Data compression, or source coding, is a scheme to 
reduce the size of message (data) in information representation. 
In his seminal paper \cite{Shannon}, Shannon showed that 
for an information source represented by a distribution 
${\cal P}(\boldsymbol{S})$ of $N$ dimensional Boolean (binary) vector 
$\boldsymbol{S}$, one can employ another representation 
in which the message length $N$ is reduced to $M(\le N)$ 
without any distortion,  if the code rate $R=M/N$ satisfies 
$R \ge H_2 \left ( \boldsymbol{S} \right )$ 
in the limit $N, \ M \to \infty$. Here,
$H_2 \left ( \boldsymbol{S} \right )=-(1/N)
\mathop{\rm Tr}_{\boldsymbol{S}} {\cal P}(\boldsymbol{S})
\log_2 {\cal P}(\boldsymbol{S})$ represents the binary 
entropy per bit in the original representation 
$\boldsymbol{S}$ indicating the optimal compression rate.

Unfortunately, Shannon's theorem itself is non-constructive and does not 
provide explicit rules for devising the optimal codes. 
Therefore, it is surprising that a practical 
code proposed by Lempel and Ziv (LZ) in 1973 \cite{Ziv} 
saturates the Shannon's optimal compression limit 
in the case of point-to-point communication. 
However, it should be emphasized here that 
generalization of the LZ codes to advanced data 
compression suitable for network communications
(NC) is difficult although importance of the NC is rapidly 
increasing as recent development of the Internet. 
This is mainly because all 
the practical codes that saturate Shannon's limit to date 
require a complete knowledge about all source
vectors coming into the communication network
while the compression should be carried out independently 
on each terminal in usual situations. 
Therefore, the quest for more efficient compression codes that are 
suitable for NC still remains one of the most important topics 
in information theory \cite{Cover}. 

The purpose of this letter is to employ recent developments 
of the research on error-correcting codes (ECC) for this purpose. 
More specifically, we will investigate the efficacy and the limitation 
of a linear compression scheme inspired by Gallager's 
ECC \cite{Gallager}, which has been 
actively investigates in both of information theory
and physics communities \cite{MacKay,Richardson,Murayama,Montanari},  
when it is applied to a data compression problem 
introduced by Slepian and 
Wolf (SW) in the research of NC \cite{Slepian}. 
Unlike the existing argument in information theory, 
our approach based on statistical mechanics
makes it possible not only to assess the theoretical bounds of 
the achievable performance but also to provide practical 
encoding/decoding methods that can be performed
in linear time scales with respect to the data length.

\begin{figure}
  \centering
  \subfigure[Slepian and Wolf system]{\label{fig:1:a}\includegraphics[scale=0.95]{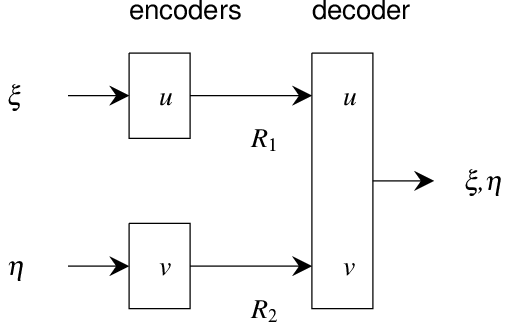}}
  \subfigure[Achievable rate region]{\label{fig:1:b}\includegraphics[scale=0.95]{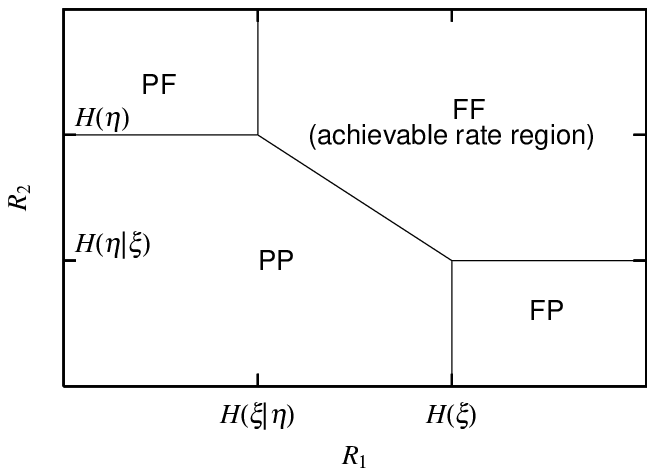}}
  \caption{(a)Slepian and Wolf system: 
A simple communication network introduced in the data compression 
theorem of Slepian and Wolf. 
Separate coding is assumed in the distributed system. 
(b) Achievable rate region: 
Code rates are classified into four categories according to 
whether the two compressed data are decodable or not. 
The parameter regime where the both data are decodable 
without any distortion is termed the {\em achievable rate region.} }
  \label{fig:1}
\end{figure}

Let us start with setting up the framework of the SW problem \cite{Slepian}. 
In a general scenario, two correlated $N$-dimensional Boolean vectors
$\boldsymbol{\xi}$ and $\boldsymbol{\eta}$ are {\em independently }
compressed to $M$-dimensional vectors $\boldsymbol{x}$ and $\boldsymbol{y}$ 
respectively. 
These compressed data (or codewords) $\boldsymbol{x}$ and 
$\boldsymbol{y}$ are decoded to retrieve the original data 
simultaneously by a single decoder. 
A schematic representation of this system is shown in Fig. \ref{fig:1:a}.

The codes used in this letter are composed of 
randomly selected sparse matrices $A$ and $B$ of 
dimensionality $M_1 \times N$ and $M_2 \times N$, respectively. 
These are constructed similarly to those of Gallager's 
ECC \cite{Gallager} as being characterized by $K_1$ and $K_2$ 
nonzero unit elements per row and $C_1$ and $C_2$ nonzero unit 
elements per column, respectively. 
The compression rates can be different between the two terminals. 
Corresponding to matrices $A$ and $B$, the rates are defined as 
$R_1=M_1/N=K_1/C_1$ and $R_2=M_2/N=K_2/C_2$, respectively. 
While both matrices are known to decoder, 
encoders only need to know their own matrix, 
that is, encoding carried out separately in this scheme
as $\bu=A \boldsymbol{\xi}$ and $\bv=B \boldsymbol{\eta}$, 
where Boolean arithmetics are employed to the Boolean vectors.
After receiving the codewords $\bu$ and 
$\bv$, the couple of equations
\begin{eqnarray}
\bu=A \boldsymbol{S}, \quad \bv=B \boldsymbol{\tau}
\label{eq:decoding}
\end{eqnarray}
should be solved with respect to $\bS$ and $\btau$ which 
become the estimates of the original data $\bxi$ and $\be$, 
respectively. 

To facilitate the current investigation we first map the problem 
to that of an Ising model with finite connectivity. We employ the 
binary representation $(+1,-1)$ of the dynamical variables 
$\boldsymbol{S}$ and $\boldsymbol{\tau}$ and of the vectors 
$\bu$ and $\bv$ rather than the Boolean $(0,1)$ one; 
the vector $\boldsymbol{u}$ is generated 
by taking products of the relevant binary data bits 
$u_{\langle i_1,i_2,\cdots,i_{K_1} \rangle}=\xi_{i_1}\xi_{i_2}\cdots 
\xi_{i_{K_1}}$, where the indices $i_1,i_2,\cdots,i_{K_1}$ 
correspond to the nonzero elements of $A$, producing a binary 
version of $\bu$, and similarly for $\bv$. 
Assuming the thermodynamic limit $N, \ M_1, \ M_2 \to \infty$ keeping 
the code rates $R_1=M_1/N $ and $R_2=M_2/N$ finite 
is quite natural as 
communication to date generally requires transmitting large data, 
where finite size corrections are likely to be negligible. 
To explore the system's capabilities we examine the partition function
\begin{eqnarray}
\mathcal{Z}=\underset{\boldsymbol{S},\boldsymbol{\tau}}{\mathrm{Tr}} 
\mathcal{P}(\boldsymbol{S},\boldsymbol{\tau}) 
\prod_{\langle i_1,i_2,\cdots,i_{K_1} \rangle} 
\left[ 1+ \frac{1}{2} \mathcal{A}_{\langle i_1,i_2,\cdots,i_{K_1} \rangle} 
\left(u_{\langle i_1,i_2,\cdots,i_{K_1} \rangle} 
\cdot S_{i_1}S_{i_2}\cdots S_{i_{K_1}}-1 \right) \right] \nonumber \\
\times
\prod_{\langle i_1,i_2,\cdots,i_{K_2} \rangle} 
\left[ 1+ \frac{1}{2} \mathcal{B}_{\langle i_1,i_2,
\cdots,i_{K_2} \rangle} \left(v_{\langle i_1,i_2,\cdots,i_{K_2} 
\rangle} \cdot \tau_{i_1}\tau_{i_2}\cdots \tau_{i_{K_2}}-1 \right) \right]. 
\label{eq:partition}
\end{eqnarray}
The tensor product $\mathcal{A}_{\langle i_1,i_2,\cdots,i_{K_1} \rangle} 
u_{\langle i_1,i_2,\cdots,i_{K_1} \rangle}$, 
where $u_{\langle i_1,i_2,\cdots,i_{K_1} \rangle}=
\xi_{i_1}\xi_{i_2}\cdots \xi_{i_{K_1}}$ is the binary 
equivalent of $A \boldsymbol{\xi}$. Elements of the sparse 
connectivity tensor $\mathcal{A}_{\langle i_1,i_2,\cdots,i_{K_1} 
\rangle}$ take the value $1$ if the corresponding indices 
of data are chosen (i.e., if all corresponding indices of the 
matrix $A$ are $1$) and $0$ otherwise; it has $C_1$ unit 
elements per $i$ index representing the system's degree of 
connectivity. Notice that if the product 
$S_{i_1}S_{i_2}\cdots S_{i_{K_1}}$ is in disagreement with 
the corresponding element $u_{\langle i_1,i_2,\cdots,i_{K_1}\rangle}$, 
which implies an error for the parity check,
the value of the partition function $\mathcal{Z}$ vanishes. 
Similar arguments are valid 
for $\mathcal{B}_{\langle i_1,i_2,\cdots,i_{K_2} \rangle}$ and 
$v_{\langle i_1,i_2,\cdots,i_{K_2} \rangle}$. 
The probability $\mathcal{P}(\boldsymbol{S},\boldsymbol{\tau})$ 
represents our prior knowledge of data 
including the correlation 
between the sources $\bxi$ and $\be$. 
Note that the dynamical variables 
$\boldsymbol{\tau}$, introduced to estimate 
$\boldsymbol{\eta}$, are irrelevant to the performance 
measure with respect to the other data $\bxi$.


Since the partition function Eq. (\ref{eq:partition}) 
is invariant under the transformations
$S_i \to S_i \xi_i$, $\tau_i \to \tau_i \eta_i$, 
$u_{\langle i_1,i_2,\cdots,i_{K_1} \rangle} \to 
u_{\langle i_1,i_2,\cdots,i_{K_1} \rangle} 
\xi_{i_1}\xi_{i_2}\cdots \xi_{i_{K_1}}=1$ and 
$v_{\langle i_1,i_2,\cdots,i_{K_2} \rangle} \to 
v_{\langle i_1,i_2,\cdots,i_{K_2} \rangle} 
\tau_{i_1}\tau_{i_2}\cdots \tau_{i_{K_2}}$ $=1$,
it is useful to decouple the correlations between 
the vectors ${\boldsymbol{S}}$, ${\boldsymbol{\tau}}$ 
and ${\boldsymbol{\xi}}$, ${\boldsymbol{\eta}}$. 
Rewriting Eq. (\ref{eq:partition}) using this gauge, 
one obtains a similar expression apart from the first 
factor which becomes 
${\mathcal{P}}({\boldsymbol{S}}\otimes 
{\boldsymbol{\xi}},{\boldsymbol{\tau}}\otimes {\boldsymbol{\eta}})$, 
where ${\boldsymbol{S}}\otimes {\boldsymbol{\xi}}=\left (S_i \xi_i \right )$ 
and ${\boldsymbol{\tau}}\otimes {\boldsymbol{\eta}}
=\left ( \tau_i \eta_i \right )$ for $i=1,2,\ldots,N$. 

The random selection of elements in ${\mathcal{A}}$ and ${\mathcal{B}}$ 
indroduces disorder to the system; 
we average the logarithm of the partition function 
${\mathcal{Z}}({\mathcal{A}},{\mathcal{B}},
{\bu},{\bv})$ 
over the disorder and the statistical properties 
of both data, using the replica method \cite{Wong}. 
In the calculation, a set of order 
parameters
$q_{\alpha,\beta,\cdots,\gamma} = 
\frac{1}{N}\sum_{i=1}^N Z_i 
S_i^{\alpha}S_i^{\beta}\cdots S_i^{\gamma}$ and 
$r_{\alpha,\beta,\cdots,\gamma} = 
\frac{1}{N}\sum_{i=1}^N Y_i 
\tau_i^{\alpha}\tau_i^{\beta}\cdots \tau_i^{\gamma} $
arise, where $\alpha,\beta,\cdots,\gamma$ 
represent replica indices, and the variables $Z_i$ and $Y_i$ 
come from enforcing the restriction of $C_1$ and $C_2$ 
connections per index, respectively as in \cite{Murayama}.

To proceed further, we have to make an assumption 
about the order parameters' symmetry. The assumption 
made here, and validated later on, 
is that of replica 
symmetry in the following representation of the order 
parameters and the related conjugate variables:
\begin{eqnarray}
q_{\alpha,\beta,\cdots,\gamma} &=& a_q \int dx \, \pi(x)x^l, 
\quad {\widehat{q}}_{\alpha,\beta,\cdots,\gamma} = a_{\widehat{q}} \int d{\widehat{x}} 
\, {\widehat{\pi}}({\widehat{x}}){\widehat{x}}^l, \nonumber \\
r_{\alpha,\beta,\cdots,\gamma} &=& a_r \int dy \, \rho(y)y^l, 
\quad {\widehat{r}}_{\alpha,\beta,\cdots,\gamma} = 
a_{\widehat{r}} \int d{\widehat{y}} \, {\widehat{\rho}}({\widehat{y}}){\widehat{y}}^l,
\end{eqnarray}
where $l$ is the number of replica indices and $a_*$ are normalization 
factors to make $\pi(x)$, ${\widehat{\pi}}({\widehat{x}})$, 
$\rho(y)$ and ${\widehat{\rho}}({\widehat{y}})$ represent 
probability distributions. Unspecified integrals are carried out 
over the range $[-1,+1]$. 

Extremizing the averaged expression 
with respect to the probability distributions, 
we obtain the following free energy per spin:
\begin{eqnarray}
{\cal F} &=& - \frac{1}{N} \langle \ln 
{\mathcal{Z}} \rangle_{{\mathcal{A}},{\mathcal{B}},{\mathcal{P}}} \nonumber \\
&=& - \underset{\pi,{\widehat{\pi}},\rho,{\widehat{\rho}}}{\mathrm{Extr}} \left\{ \frac{C_1}{K_1} \left\langle \ln \left( \frac{1+\prod_{i=1}^{K_1}x_i}{2} \right) \right\rangle_{\pi}+\frac{C_2}{K_2} \left\langle \ln \left( \frac{1+\prod_{i=1}^{K_2}y_i}{2} \right) \right\rangle_{\rho} \right. \nonumber \\
&\phantom{=}& \left. -C_1 \left\langle \ln \left( \frac{1+x {\widehat{x}}}{2} \right) \right\rangle_{\pi,{\widehat{\pi}}}-C_2 \left\langle \ln \left( \frac{1+y {\widehat{y}}}{2} \right) \right\rangle_{\rho,{\widehat{\rho}}} \right. \nonumber \\
&\phantom{=}& \left. +\frac{1}{N} \left\langle \ln \left[ \underset{{\boldsymbol{S}},{\boldsymbol{\tau}}}{{\mathrm{Tr}}} \prod_{i=1}^N \prod_{\mu=1}^{C_1} \left( \frac{1+{\widehat{x}}_{\mu i} S_i}{2} \right) \prod_{i=1}^N \prod_{\mu=1}^{C_2} \left( \frac{1+{\widehat{y}}_{\mu i} \tau_i}{2} \right) {\mathcal{P}}({\boldsymbol{S}}\otimes {\boldsymbol{\xi}},{\boldsymbol{\tau}}\otimes {\boldsymbol{\eta}}) \right] \right\rangle_{{\widehat{\pi}},{\widehat{\rho}},{\mathcal{P}}} \right\},
\label{eq:free_energy}
\end{eqnarray}
where 
the brackets with the subscript $\pi$ and ${\widehat{\pi}}$ 
represent averages over the probability distributions
$\pi(x)$ and $\pi (\widehat{x})$ with respect to 
variables denoted by $x$ and $\widehat{x}$ with and without 
subscripts, respectively.
Similar notations are also used for $\rho$ and ${\widehat{\rho}}$.
The bracket with the subscript ${\mathcal{P}}$ denotes 
the average with respect to $\bxi$ and $\be$ following 
the data distribution ${\cal P}(\bxi,\be)$. 

Taking the functional derivative with respect to the distributions 
$\pi$, ${\widehat{\pi}}$, 
$\rho$ and ${\widehat{\rho}}$, we obtain the following saddle point equations:
\begin{eqnarray}
\pi(x) &=& \frac{1}{N} \sum_{i=1}^N \left\langle 
\delta \left[ x - \tanh \left( F_i({\widehat{x}}_{\mu j \in {\mathcal{L}}(\mu)/i}, 
{\widehat{y}}_{\mu i};{\boldsymbol{\xi}},{\boldsymbol{\eta}}) 
\xi_i + \sum_{\mu=1}^{C_1-1} \tanh^{-1}({\widehat{x}}_{\mu i}) 
\right) \right] \right\rangle_{{\widehat{\pi}},{\widehat{\rho}},{\mathcal{P}}}
\nonumber \\
{\widehat{\pi}}({\widehat{x}}) &=& \left\langle \delta \left( {\widehat{x}} - \prod_{i=1}^{K_1-1} x_i \right) \right\rangle_{\pi} \label{eq:saddle}
\end{eqnarray}
where the effective fields denoted by $F_i$ with 
subscripts are implicitly defined as 
\begin{eqnarray}
&\phantom{=}& \frac{e^{F_i({\widehat{x}}_{\mu j \in {\mathcal{L}}(\mu)/i}, {\widehat{y}}_{\mu i};{\boldsymbol{\xi}},{\boldsymbol{\eta}}) \xi_i S_i}}{2 \cosh F_i({\widehat{x}}_{\mu j \in {\mathcal{L}}(\mu)/i}, {\widehat{y}}_{\mu i};{\boldsymbol{\xi}},{\boldsymbol{\eta}})} \nonumber \\
&=& \frac{\underset{{\boldsymbol{S}}/S_i,{\boldsymbol{\tau}}}{\mathrm{Tr}} \prod_{j \in {\mathcal{L}}(\mu)/i} \prod_{\mu=1}^{C_1} \left( \frac{1+{\widehat{x}}_{\mu j} S_j}{2} \right) \prod_{i=1}^N \prod_{\mu=1}^{C_2} \left( \frac{1+{\widehat{y}}_{\mu i} \tau_i}{2} \right) {\mathcal{P}}({\boldsymbol{S}}\otimes {\boldsymbol{\xi}},{\boldsymbol{\tau}}\otimes {\boldsymbol{\eta}})}{\underset{{\boldsymbol{S}},{\boldsymbol{\tau}}}{\mathrm{Tr}} \prod_i^N \prod_{\mu=1}^{C_1} \left( \frac{1+{\widehat{x}}_{\mu i} S_i}{2} \right) \prod_{i=1}^N \prod_{\mu=1}^{C_2} \left( \frac{1+{\widehat{y}}_{\mu i} \tau_i}{2} \right) {\mathcal{P}}({\boldsymbol{S}}\otimes {\boldsymbol{\xi}},{\boldsymbol{\tau}}\otimes {\boldsymbol{\eta}})},
\end{eqnarray}
and similarly for $\rho(y)$ and ${\widehat{\rho}({\widehat{y}})}$. Notice that the notation ${\boldsymbol{S}}/S_i$ represents the set of all dynamical variables ${\boldsymbol{S}}$ except $S_i$. On the other hand, ${\mathcal{L}}_1(\mu)$ 
and ${\mathcal{L}}_2(\mu)$ denote the set of all indices 
of nonzero components in the $\mu$th row of $A$ and $B$, respectively. 
The notation ${\mathcal{L}}_1(\mu)/i$ represents 
the set of all indices belonging to ${\mathcal{L}}_1(\mu)$ 
except $i$, and similarly for others. 

After solving these equations, the expectation of the 
overlap can be evaluated as
\begin{eqnarray}
{\mathsf{m}}_1 = \frac{1}{N} \left\langle \sum_{i=1}^N \xi_i 
\, {\mathrm{sign}} \, \langle S_i \rangle 
\right\rangle_{\mathcal{A},\mathcal{P}} = 
\int dz \, \phi (z) \, {\mathrm{sign}} (z),
\end{eqnarray}
where we denote thermal averages $\langle \cdots \rangle$ and
\begin{eqnarray}
\phi(z)=\frac{1}{N}\sum_{i=1}^N 
\left\langle \delta \left[ z-\tanh \left( F_i({\widehat{x}}_{\mu j \in {\mathcal{L}}(\mu)/i}, 
{\widehat{y}}_{\mu i};{\boldsymbol{\xi}},{\boldsymbol{\eta}})\xi_i 
+\sum_{i=1}^{C_1} \tanh^{-1} {\widehat{x}}_i \right) \right] 
\right\rangle_{{\widehat{\pi}},{\widehat{\rho}}, {\mathcal{P}}},
\end{eqnarray}
and similarly for ${\mathsf{m}}_2$ of the overlap between 
${\boldsymbol{\eta}}$ and its estimator. 

The performance of the current compression method can be 
measured by the vector $\boldsymbol{\mathsf{m}}=({\mathsf{m}}_1,{\mathsf{m}}_2)$. 
Hereafter, we use the term `ferromagnetic' to specify 
the perfect retrieval, that is , ${\mathsf{m}}_1=1$ (or ${\mathsf{m}}_2=1$), while  the term `paramagnetic' implies the distortion, that is, 
${\mathsf{m}}_1 < 1$ (or ${\mathsf{m}}_2 < 1$). For instance, a term such as 
`ferromagnetic-paramagnetic phase' denotes the phase characterized 
by the performance vector 
${\boldsymbol{\mathsf{m}}} \in \{ ({\mathsf{m}}_1,{\mathsf{m}}_2) | 
{\mathsf{m}}_1=1, {\mathsf{m}}_2<1 \}$, and so on.

One can show that the ferromagnetic-ferromagnetic state (FF):
$\pi(x) = \delta (x-1), \ {\widehat{\pi}}({\widehat{x}}) = 
\delta ({\widehat{x}}-1), \ 
\rho(y) = \delta (y-1)$ and 
${\widehat{\rho}}({\widehat{y}}) = \delta ({\widehat{y}}-1)$
always satisfies Eq. (\ref{eq:saddle}). 
In addition, in the limit of $C_1, \ C_2 \to \infty$, 
four solutions describing
the paramagnetic-paramagnetic state (PP):
$\pi(x) = \delta (x), \ {\widehat{\pi}}({\widehat{x}}) = \delta ({\widehat{x}}), 
\ \rho(y) = \delta (y)$ and ${\widehat{\rho}}({\widehat{y}}) = \delta ({\widehat{y}})$, 
the paramagnetic-ferromagnetic phase (PF):
$\pi(x) = \delta (x), \ {\widehat{\pi}}({\widehat{x}}) 
= \delta ({\widehat{x}}), \ 
\rho(y) = \delta (y-1)$ and ${\widehat{\rho}}({\widehat{y}}) = 
\delta ({\widehat{y}}-1)$ and the ferromagnetic-paramagnetic state (FP):
$\pi(x) = \delta (x-1), \ {\widehat{\pi}}({\widehat{x}}) 
= \delta ({\widehat{x}}-1), \ 
\rho(y) = \delta (y)$ and 
$\widehat{\rho}({\widehat{y}})= \delta(\widehat{y})$
are also analytically obtained for an {\em arbitrary} joint 
distribution ${\cal P}(\bxi,\be)$. 
Free energies corresponding to these solutions are 
provided from Eq. (\ref{eq:free_energy}) as 
\begin{eqnarray}
{\cal F}_{FF}&=&-\frac{1}{N} \mathop{\rm Tr}_{\bxi,\be} 
{\cal P}(\bxi,\be) \ln {\cal P}(\bxi,\be), \
{\cal F}_{PP} = (R_1+R_2) \ln 2, \cr 
{\cal F}_{FP} &=& R_2 \ln 2 -
\frac{1}{N} \mathop{\rm Tr}_{\bxi} {\cal P}(\bxi) \ln 
{\cal P}(\bxi), \
{\cal F}_{PF} = R_1 \ln 2 -
\frac{1}{N} \mathop{\rm Tr}_{\be} {\cal P}(\be) \ln {\cal P}(\be), 
\label{eq:free_energies}
\end{eqnarray}
where subscripts stand for corresponding states and 
${\cal P}(\bxi)=\mathop{\rm Tr}_{\be} {\cal P}(\bxi, \be) $
and ${\cal P}(\be)=\mathop{\rm Tr}_{\bxi} {\cal P}(\bxi,\be)$
represent marginal distributions for the two 
source vectors $\bxi$ and $\be$, respectively. 

Perfect decoding is theoretically possible 
if ${\cal F}_{FF}$ is the lowest among the above four. 
The corresponding parameter regime termed 
{\em achievable rate region} is shown in Fig. \ref{fig:1:b} as 
an intersection of the inequalities
\begin{eqnarray}
R_1+R_2 \ge H_2({\boldsymbol{\xi}},{\boldsymbol{\eta}}), 
\quad R_1 \ge H_2({\boldsymbol{\xi}}|{\boldsymbol{\eta}}), 
\quad R_2 \ge H_2({\boldsymbol{\eta}}|{\boldsymbol{\xi}}),
\end{eqnarray}
where $H_2({\boldsymbol{\xi}},{\boldsymbol{\eta}})
=-(1/N)\mathop{\rm Tr}_{\bxi,\be}
{\cal P}(\bxi,\be)\ln {\cal P}(\bxi,\be)$, 
$H_2({\boldsymbol{\xi}}|{\boldsymbol{\eta}})
=H_2(\bxi,\be)-H_2(\be)$ and 
$H_2({\be}|{\bxi})
=H_2(\bxi,\be)-H_2(\bxi)$. 
It is worthy of noticing that this coincides 
with the achievable rate region saturated by the {\em optimal }
data compression in the current framework previously shown by 
SW \cite{Slepian}. 
Namely, in the limit $C_1, \ C_2 \to \infty$, the current compression 
codes provide the optimal performance for {\em arbitrary} 
information sources ${\cal P}(\bxi,\be)$.

For finite $C_1$ and $C_2$, the saddle point equations (\ref{eq:saddle}) 
can be solved numerically; but the properties of the system highly 
depend on the source distribution ${\cal P}(\bxi,\be)$, 
which makes it difficult to go further without any assumption 
on the distribution. 
As a simple but non-trivial example, we will focus here
on a component-wise correlated joint distribution
\begin{eqnarray}
{\mathcal{P}}({\boldsymbol{S}},{\boldsymbol{\tau}})=
\prod_{i=1}^N \left( \frac{1+m_1 S_i+m_2 \tau_i +q S_i \tau_i}{4} 
\right), 
\label{eq:correlation}
\end{eqnarray}
where a set of parameters $m_1$, $m_2$, and $q$ characterize 
the data sources. To make Eq. (\ref{eq:correlation}) a distribution, 
these parameters must satisfy four inequalities
$1+m_1+m_2+q \ge 0$, $1-m_1+m_2-q \ge 0$, $1+m_1-m_2-q \ge 0$
and $1-m_1-m_2+q \ge 0$. 

Solving Eq. (\ref{eq:decoding}) rigorously for decoding 
is computationally hard in general cases. 
However, one can construct a practical decoding algorithm 
based on the belief propagation (BP) \cite{Pearl} 
or the Thouless-Anderson-Palmer (TAP)
approach \cite{TAP}. It has recently been shown that these two frameworks 
provide the same algorithm in the case of ECC \cite{Kabashima}. 
This is also the case under the current context. 
For distribution (\ref{eq:correlation}), the algorithm derived 
from the BP/TAP frameworks becomes as 
\begin{eqnarray}
m_{\mu i}^1 &=& \frac{a_{\mu i}+ m_1 + m_2 a_{\mu i} b_i 
+ q b_i}{1+m_1 a_{\mu i} +m_2 b_i+q a_{\mu i} b_i}, 
\quad m_{\mu i}^2 = \frac{b_{\mu i}+ m_2 + m_1 a_i b_{\mu i} 
+ q a_i}{1+m_1 a_i +m_2 b_{\mu i}+q a_i b_{\mu i}}, \nonumber \\
{\widehat{m}}_{\mu i}^1 &=& u_{\mu} \prod_{j \in {\mathcal{L}}_1(\mu)/i} 
m_{\mu j}^1, \quad {\widehat{m}}_{\mu i}^2 = 
v_{\mu} \prod_{j \in {\mathcal{L}}_2 (\mu)/i} m_{\mu j}^2, 
\label{eq:belief}
\end{eqnarray}
where we denote $a_{\mu i} \equiv \tanh 
\sum_{\nu \in {\mathcal{M}}_1(i)/\mu} \tanh^{-1} 
{\widehat{m}}_{\nu i}^1$ and $a_i \equiv 
\tanh \sum_{\mu \in {\mathcal{M}}_1(i)} 
\tanh^{-1} {\widehat{m}}_{\mu i}^1$, and 
similarly for $b$'s. 
Here, ${\mathcal{M}}_1(i)$ and ${\mathcal{M}}_2(i)$ 
indicate the set of all indices of nonzero components 
in the $i$th column of the sparse matrices $A$ and $B$, 
respectively. 
Eq. (\ref{eq:belief}) can be solved iteratively 
from the appropriate initial conditions. 
After obtaining a solution, approximated posterior means 
can be calculated for $i=1,2,\ldots,N$ as
\begin{eqnarray}
m_i^1 = \left \langle S_i \right \rangle =\frac{a_i+ m_1 + m_2 a_i b_i + q b_i}
{1+m_1 a_i +m_2 b_i+q a_i b_i}, 
\quad m_i^2 = \left \langle \tau_i \right \rangle =
\frac{b_i+ m_2 + m_1 a_i b_i + q a_i}
{1+m_1 a_i +m_2 b_i+q a_i b_i},
\end{eqnarray}
which provide an approximation to the Bayes-optimal 
estimators as 
$\xi_i={\mathrm{sign}}(m_i^1)$ and 
$\eta_i={\mathrm{sign}}(m_i^2)$, respectively.

\begin{figure}
 \begin{center}
 \includegraphics[scale=0.95]{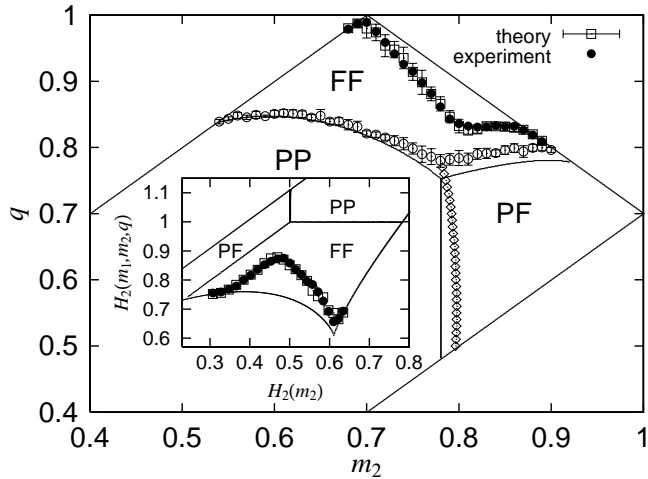}
 \end{center}
 \caption{Phase diagram for $K_1=K_2=6$ and $C_1=C_2=3$ code in the case of component-wise correlated information source (\ref{eq:correlation}). This figure 
shows that the feasible region in $m_2-q$ plane for $m_1=0.7$ is 
classified into three states. Phase boundaries obtained by 
numerical methods are indicated by $\circ$ 
with errorbars (FF/PP and FF/PF) and $\Diamond$ (PF/PP). 
These are close to those for $K_1=K_2 \to \infty, C_2=C_2 
\to \infty$ (curves and the vertical line). 
Practically decodable limits of the TAP/BP algorithm 
obtained for $N=10^4$ systems are indicated as $\bullet$. 
These are well evaluated by the spinodal 
points of non-FF solutions ($\Box$ with errorbars). 
Inset: The practical limits are represented 
by the sizes of transmitted information. 
The horizontal and vertical axes show the entropy of the second 
source ${\boldsymbol{\tau}}$ and the joint entropy, respectively. 
}
 \label{fig:2}
\end{figure}

In order to investigate the efficacy of the current 
method for finite $C_1$ and $C_2$, we have numerically 
solved Eqs. (\ref{eq:saddle}) and (\ref{eq:belief})
for $K_1=K_2=6$ and $C_1=C_2=3$ $(R_1=R_2=1/2)$, 
results of which are summarized in Fig. \ref{fig:2}. 
Numerical results for the saddle point equation (\ref{eq:saddle}) 
were obtained by an iterative method using $10^4-10^5$ bin models 
for each probability distribution.
$10-10^2$ updates were sufficient for convergence in most cases. 
Similarly to the case of $C_1, \ C_2 \to \infty$, there can be 
four types of solutions corresponding to combinations 
of decoding success and failure on the two sources. 
The obtained phase diagram is quite similar to that for 
$C_1, \ C_2 \to \infty$. 
This implies that 
the current compression code {\em theoretically }
has a good performance close to the optimal one that is saturated in 
the limit $C_1, \ C_2 \to \infty$ although 
the choice of $C_1 =C_2=3$ is far from such limit. 

However, this does not directly mean that the suggested performance can be 
obtained {\em in practice}. Since the variables are updated locally in 
the BP/TAP decoding algorithm (\ref{eq:belief}), it may become 
difficult to find the thermodynamically dominant state when 
there appear suboptimal states which have large basins of attraction. 
This suggests that the practical performance for the perfect decoding 
is determined by the spinodal points of the suboptimal states
similarly to the case of ECC \cite{Murayama}. 
In order to confirm this conjecture, we have numerically compared
the practical limit of the perfect decoding 
obtained by the BP/TAP decoding algorithm (\ref{eq:belief}) and 
the spinodal points of the non-FF solutions. 
These two results exhibit an excellent consistency 
supporting our conjecture. 
In the figure, the perfectly decodable region 
obtained by the BP/TAP algorithm for $m_1=0.7$ cases 
is indicated as the area surrounded 
by the spinodal points and the boundaries for the feasible region 
$1+0.7-m_2-q=0$ and $1+0.7+m_2-q=0$. 
This looks narrow compared to the theoretical limit, which 
might provide a negative impression on the practical utility  
of this code. Nevertheless, we still consider that the current 
method may be practically useful because the size of information 
that can be represented by parameters in the region 
is not so small as what the area looks. 

In summary, we have developed an efficient method of 
data compression in a 
multi-terminal communication network, 
taking advantage of the 
sparse matrix based linear compression codes. 
We observed several practical properties of the codes of this type 
in the simplest model of a data compression 
employed for a network communication proposed by SW. 
Studying the typical performance of the linear 
compression codes in a network, which complements 
the methods used in the information theory literature, 
is the first step towards understanding typical properties 
of the network based systems.  


\acknowledgments
The author thanks Y. Kabashima and T. Ohira for valuable discussions 
and suggestions. This work was partly supported by 
the Japan Society for the Promotion of Science.

\end{document}